# Lesion classification by model-based feature extraction: A differential affine invariant model of soft tissue elasticity

Weiguo Cao [1], Marc J. Pomeroy [1,2], Zhengrong Liang [1,2,*], Yongfeng Gao [1], Yongyi Shi [1], Jiaxing Tan [3], Fangfang Han [4], Jing Wang [5], Jianhua Ma [4], Hongbin Lu [6], Almas F. Abbasi [1], and Perry J. Pickhardt [7]

[1] Department of Radiology, Stony Brook University, Stony Brook, NY 11794, USA.
[2] Department of Biomedical Engineering, Stony Brook University, Stony Brook, NY 11794, USA
[3] Department of Computer Science, City University of New York at CSI, NY 10314, USA.
[4] School of Biomedical Engineering, Southern Medical University, Guangzhou, Guangdong, China.
[5] Department of Radiation Oncology, University of Texas Southwestern Medical Centre, Dallas, TX 75235, USA.
[6] Department of Biomedical Engineering, The Fourth Medical University, Xi'an, China.
[7] Department of Radiology, School of Medicine, University of Wisconsin, Madison, WI 53792, USA.
[*] Corresponding author (e-mail: jerome.liang@sunysb.edu)

___________________________________________________________________________________________

**Abstract.** The elasticity of soft tissues has been widely considered as a characteristic property to differentiate between healthy and vicious tissues and, therefore, motivated several elasticity imaging modalities, such as Ultrasound Elastography, Magnetic Resonance Elastography, and Optical Coherence Elastography. This paper proposes an alternative approach of modeling the elasticity using Computed Tomography (CT) imaging modality for model-based feature extraction machine learning (ML) differentiation of lesions. The model describes a dynamic non-rigid (or elastic) deformation in differential manifold to mimic the soft tissues' elasticity under wave fluctuation *in vivo*. Based on the model, three local deformation invariants are constructed by two tensors defined by the 1$^{st}$ and 2$^{nd}$ order derivatives from the CT images and used to generate elastic feature maps after normalization via a novel signal suppression method. The model-based elastic image features are extracted from the feature maps and fed to machine learning to perform lesion classifications. Two pathologically proven image datasets of colon polyps (44 malignant and 43 benign) and lung nodules (46 malignant and 20 benign) were used to evaluate the proposed model-based lesion classification. The outcomes of this modeling approach reached the score of area under the curve of the receiver operating characteristics of 94.2% for the polyps and 87.4% for the nodules, resulting in an average gain of 5% to 30% over ten existing state-of-the-art lesion classification methods. The gains by modeling tissue elasticity for ML differentiation of lesions are striking, indicating the great potential of exploring the modeling strategy to other tissue properties for ML differentiation of lesions.

___________________________________________________________________________________________

## 1. Introduction

Lesion classification plays a critical role in relieving the concerns on over-detection (high false positive rate) and under-diagnosis (low accuracy) in medical imaging-based lesion characterization. Conventional machine learning (cML) extracts abstract image features based on some visible image contrast patterns, such as image textures, followed by a feature classifier for lesion diagnosis. Recent deep learning (DL) extracts abstract image features and classifies the features simultaneously by a self-adjustable manner in the design of a single architecture. Since the visible image contrast patterns of the lesion volume represent mainly the distributions of the image gray levels across the locations of all image elements or voxels, the extracted abstract image features reflect mainly the static properties of the tissues inside the lesion volume. Thus, both cML and DL are lacking the capability of fully considering the tissue properties in the feature extraction stage, particularly the tissue dynamic properties, which are mainly reflected by the contrast changing at each voxel, including the changing direction, the changing rate, etc. Modeling strategy reflects human intelligence in problem solving in various conditions [1], including limited dataset condition which is common in medical lesion classification. This study adapts the strategy to model the tissue properties for feature extraction in order to predict the lesion malignancy. To demonstrate the potential of tissue modeling strategy for lesion classification, this study takes the tissue elasticity property as an example.

The elasticity of *in vivo* soft tissues has been proved to bring discriminative information for use in diagnosing pathologic changes of tissues in lesions [2,3]. The elastic modulus (ESM) has been recognized as one of the most important metrics of the elasticity [4]. The ESM is usually expressed by the ratio of stress and strain in the elastic deformation region, where the stress is the unit force causing deformation and the strain is the ratio of the change caused by the displacement to its original shape [5]. Calculating the ESM usually requires obtaining stress and strain over a short time period, i.e. acquiring a series of images, with certain mechanical vibration equipment. Currently the ESM is calculated from a series of images reconstructed by ultrasound [5], optics [6], and magnetic resonance [7].

Intuitively, the displacement over a very short time of a continuous soft tissue elastic deformation is reflected by the tissue dynamical properties, e.g., the motion trend, which can be mathematically specified by the $1^{st}$- and $2^{nd}$-order derivatives of the image contrast distribution across the group of soft tissues inside the lesion volume. Based on this intuition, this paper proposes a model of computing the ESM image features from a single scan of computed tomography (CT) imaging modality for a model-based computer-aided diagnosis (CADx) of lesions. The model describes a dynamic elastic or non-rigid deformation in differential manifold to mimic the soft tissues' elasticity under wave fluctuation *in vivo*. Based on the model, three local deformation invariants are constructed by two tensors defined by the $1^{st}$- and $2^{nd}$-order derivatives from the CT image and used to generate ESM feature maps after normalization via a novel signal suppression method. The elastic image features can be extracted from the ESM feature maps and fed into a classifier, called adaptive ML (aML) hereafter, for CADx of the lesion. The validity of the intuition or the proposed elastic dynamic model is tested by comparing the outcomes of classifying the model-based elastic image features to the outcomes of classification of existing state-of-the-art image features from pathologically proven lesion CT image datasets of limited sizes.

## 2. Methods

In this section, we begin with the dynamic deformation model for soft tissue's motion trend by combining the differential manifold and the local affine transformation to mimic the shape deformation of the soft tissue caused by its motions or wave fluctuations *in vivo*. To describe the intrinsic properties of tissue elasticity, three new measures of the elasticity, called differential elastic measures (DEMs), are extracted from the differential domain based on the dynamic model in three-dimensional (3D) Euclidean space. To make good use of the DEMs, a signal suppression method is devised to perform signal normalization and digitalization to generate elastic feature maps of the soft tissues. The elastic feature maps are then used to construct quantitative elastic feature measures or descriptors of lesions for CADx applications.

*2.1. Dynamic Deformation Model for Soft Tissue's Motion*

Different from the traditional models of soft tissue's elasticity [4,8-10], we attempt to address this issue from another viewpoint of the shape deformations caused by the periodic motion which can be considered as a kind of wave fluctuation relevant to inner or outer forces and elasticity of the soft tissues. This shape deformation should be a type of dynamic and non-rigid deformation which is a complex problem involving physics, material sciences, geometry and approximation theory. Therefore, we address this problem under the following three assumptions.

*2.1.1. Soft tissues are 3D differential manifolds*

Soft tissues can be categorized as non-rigid objects and modeled by the differentiable manifold in the 3D Euclidean space $R^3$ [11]. For any given non-rigid object O=O($x,y,z$), we assume that it is a m-order differentiable function defined on the differentiable manifold where m ≥ 2.

*2.1.2. Complex non-rigid deformations can be decomposed into a large quantity of local shape deformations*

*In vivo* soft tissue can be treated as the composition of many micro-structures which correspond to very small local shape units. While the *in vivo* soft tissue is fluctuating, the fluctuation will produce a very complex non-rigid deformation. In the micro-structure level, the local shape associated with the local fluctuation will be concurrently changed, which means the local shape deformation happens. The local shape deformation reflects the soft tissue's local motion and could be presented as a series of local non-uniform deformations. Hence, the complex non-rigid deformation is divided into a great number of local non-uniform shape deformations. This assumption is derived from the concept of differential geometry [11] which supports the ideas of integrating a large number of local tiny deformations to mimic a global non-rigid shape deformation. Therefore, local shape deformations can be non-uniform and can describe the properties of anisotropic soft tissues and overcome the limitations of the traditional modulus models [12-15] restricted by the tissue's isotropy.

*2.1.3. The local deformation relevant to the soft tissue's elasticity can be expressed by the local affine mapping*

According to the differential concept [11], a local deformation in a very small region could be modeled by a linear transformation [16]. Compared with other linear transforms such as rotation, shift, and scaling, the affine mapping involves more parameters which could handle more complex transforms, such as stretching, shearing and flipping, etc. [17-19], see **Fig. 1**. Therefore, the local affine transformation has the capability to truly mimic most deformations caused by soft tissue motions, which could be referred to as the result of elastic wave fluctuation. Additionally, this assumption also supports that every local deformation has its individual affine parameters, where the parallelism is the only requirement in the local affine mapping. Moreover, this $3^{rd}$ assumption also implies that soft tissue deformation is restricted by boundary points.

Based on the three aforementioned assumptions, we propose a new deformation model. Let $I_1$ and $I_2$ represent two 3D volumetric image data of the same soft tissue object before and after shape deformation where coordinates ($x,y,z$) and ($X,Y,Z$) are two corresponding points in $I_1$ and $I_2$ respectively. The deformation can be modeled by the local affine mappings $P: I_1 \to I_2$ and expressed mathematically as:

$$\lim_{\substack{V_{B_k} \to 0 \\ (x,y,z) \in B_k, \\ B_k \subset I_1}} (x,y,z)^T = \lim_{\substack{V_{C_k} \to 0 \\ (X,Y,Z) \in C_k, \\ C_k \subset I_2}} P * (X,Y,Z)^T \tag{1}$$

where $B_k$ and $C_k$ are two very small local patches of $I_1$ and $I_2$, respectively, $V_{B_k}$ and $V_{C_k}$ are their local volume size, $T$ is the transpose operator, notation * denotes the matrix multiplication, and $P$ is a local affine matrix which is related to the stress and strain over the soft tissue:

$$\begin{cases} P = \begin{bmatrix} p_{11} & p_{12} & p_{13} \\ p_{21} & p_{22} & p_{23} \\ p_{31} & p_{32} & p_{33} \end{bmatrix}, |P| \neq 0 \\ p_{ij} \in \mathbb{R}, \quad \mathbb{R} \in [-\infty, +\infty], i,j = 1,2,3 \end{cases} \tag{2}$$

where $|\cdot|$ stands for the matrix determinant. Theoretically, the affine mapping $P$ between two local corresponding areas has a great number of alternatives, which compose the affine groups $\{P_1, P_2, \cdots, P_k, \cdots\}$, where $k$ is a positive integer. **Eq. (2)** also indicates that our dynamic deformation model resolves the limitation of shear modulus and bulk modulus [4] since it permits both shape changes and volume changes. Furthermore, $|P| \neq 0$ means that our dynamic model is a reversible model which guarantees the deformation recovery capability of the soft tissue under the wave fluctuation *in vivo*. $P$ must satisfy the boundary condition which forms the constraint in our dynamic model. Since the boundary is the overlap of two local neighbors, all points on it should hold the same after deformation by adjacent mapping matrix.

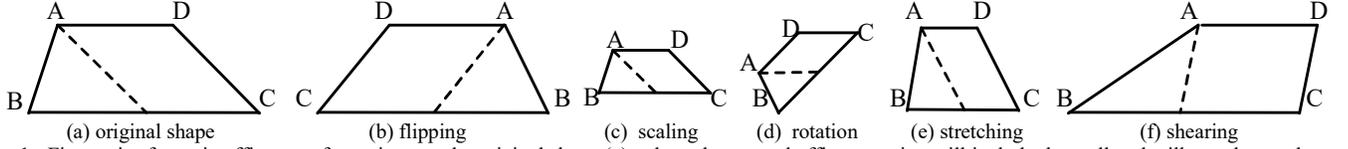

**Fig. 1**: Five major forms in affine transformations on the original shape (a), where the general affine mapping will include them all and will not change the parallelism of the original. (b) the flipping transformation of (a) which is a symmetrical form and keeps its area (2D) or volume (3D); (c) the scaling of (a) which only changes the area (2D) or volume (3D), but keeps the shape; (d) the rotation of (a) which only brings the orientation changes but keeps the shape, area (2D) or volume (3D); (e) the stretching of (a) which changes the shape, area (2D) or volume (3D), but keeps its orientation; and (f) the shearing of (a) which causes shape deformation but keeps its area (2D) or volume (3D).

According to [4], elasticity is one important intrinsic property of the soft tissue which should be independent with the shape changes and the inner/outer forces. Due to the non-isotropic nature of the soft tissue's structure, its deformations caused by the motions or wave fluctuations should contain a great number of local affine mappings despite the boundary constraints. To solve this conflict between the uncertainty of deformation parameters and the stability of its elasticity, a feasible solution is to devise some local features as the measures of its elasticity which are independent of local affine mappings $P$, despite a large number of affine transformations. Furthermore, as the displacement goes to very small toward zero, $P: I_1 \to I_2$, the local affine mappings shall be relevant to the local partial derivatives of the image $I(x, y, z)$. This is a major point of using local affine mapping to model the soft tissues' elasticity.

### 2.2. Dynamic Deformation Model for Soft Tissue's Motion

The dynamic model is constructed over small patches in 3D objects to mimic local motions or wave fluctuations of the soft tissue which are closely related to its elasticity. That means our model attempts to address the issue of the elasticity's measure in the differential domain which is also the requirement of the anisotropic soft tissue. Therefore, we study the tissue elasticity with the local shape changes and local wave fluctuations. Several differential operators are defined to describe shape deformations and wave fluctuations as follows.

Let $I = I(x, y, z)$ be a 3D image (volumetric data) of soft tissues, its gradient operator can be expressed as a vector by:

$$\nabla I = (I'_x, I'_y, I'_z) \tag{3}$$

where $I'_x, I'_y$, and $I'_z$ are the 1$^{st}$-order partial derivatives of the volumetric data of $I(x, y, z)$. In the dynamic deformation model, the gradient represents the maximum local shape changes at every point along one direction under some stressing or pushing *in vivo*. Once the local affine transformation, $P$, happens, the gradient and $P$ have the following relationship,

$$\nabla I_a = \nabla I * P \tag{4}$$

where $\nabla I_a$ is the gradient after the affine transformation.

Based on the definition of the gradient and its applications, the Harris operator, $G$, is defined as a structure tensor or matrix which is widely used to represent the full local deformation along all directions [20]:

$$G = (\nabla I)^T \cdot (\nabla I) = \begin{bmatrix} I'_x I'_x & I'_x I'_y & I'_x I'_z \\ I'_x I'_y & I'_y I'_y & I'_y I'_z \\ I'_x I'_z & I'_y I'_z & I'_z I'_z \end{bmatrix}. \tag{5}$$

**Eq. (5)** shows that the Harris operator is a symmetric matrix. We prove the following relationship between the Harris operator and the local affine transformation (details of the proof are given in the **Appendix** of this of paper).

**Property 1**

*Harris operator, $G$, after a local affine transformation keeps*

$$\mathcal{G} = P^T G P \tag{6}$$

where $\mathcal{G}$ is the subsequent Harris operator, $P$ and $P^T$ are the local affine projection and its transpose.

In addition to the Harris operator, Hessian operator is another popular tensor expressed by a squared matrix of the 2nd-order partial derivatives of a scalar-valued function [16]. Since the Laplacian operator and its trace are often utilized to express the Helmholtz equation and widely used to study the wave fluctuation [21] and elasticity calculation [22], Hessian matrix, H, not only involves shape information but also contains abundant elastic fluctuation information locally. It can be expressed as:

$$H = \begin{bmatrix} I''_{xx} & I''_{xy} & I''_{xz} \\ I''_{xy} & I''_{yy} & I''_{yz} \\ I''_{xz} & I''_{yz} & I''_{zz} \end{bmatrix} \tag{7}$$

where $I''_{xx}, I''_{xy}, I''_{xz}, I''_{yy}, I''_{yz},$ and $I''_{zz}$ are the 2nd-order partial derivatives of $I(x, y, z)$. From definitions of Hessian operator, **Eq. (7)** is also a symmetric matrix, and has the following properties under local affine transformations.

**Property 2**

*Hessian operator, $H$, after a local affine transformation keeps:*

$$\mathcal{H} = P^T H P \tag{8}$$

where $\mathcal{H}$ is the subsequent Hessian operator, $P$ is a local affine projection and $P^T$ is its transpose.

Since its proof is similar to **Property 1**, the proof of **Property 2** is omitted in this article. From **Eq. (6)** and **Eq. (8)**, the two properties give a clear relationship between the original forms and the post-mapping forms of the Harris and Hessian operators after the affine transformation. **Property 1** and **Property 2** also show that the affine mapping in our dynamic deformation model, $P$, can be extracted from the Harris and Hessian operators.

Since $H$ and $G$ are two different matrices, we use them to define two new hybrid tensors as the following:

$$\begin{cases} K_1 = H - G \\ K_2 = H + G \end{cases} \tag{9}$$

Then we continue to define two scalars based on $K_1$ and $K_2$ in differential space at non-critical points ($|H| \neq 0$),

$$\begin{cases} F_1 = \frac{|K_1|}{|H|} \\ F_2 = \frac{|K_2|}{|H|} \end{cases} \tag{10}$$

From these two properties, we have the following conclusions.

**Lemma 1**

*$F_1$ and $F_2$ are independent of the local affine mapping $P$, and they are two differential affine invariants in the 3D Euclidean space.*

The details of proof of **Lemma 1** are given in the **Appendix**, where two important scalars are defined to measure the elasticity of the soft tissues.

Similarly, we can study the determinant difference between $|K_1|$ and $|H|$ and obtain the following expression:

$$|K_1| - |H| = -(\nabla I)|H|H^{-1}(\nabla I)^T = -|H|((\nabla I)H^{-1}(\nabla I)^T) \tag{11}$$

where $H^{-1}$ is the inverse matrix of $H$ and $(\nabla I)^T$ is the transposed gradient $\nabla I$. The proof of **Eq. (11)** is given in the **Appendix**. Then we have the following equation,

$$|K_1| = |H| - |H|\left((\nabla I)H^{-1}(\nabla I)^T\right) = |H|(1 - (\nabla I)H^{-1}(\nabla I)^T) \tag{12}$$

Combining **Eq. (10)** and **Eq. (12)**, we have:

$$F_1 = \frac{|K_1|}{|H|} = \frac{|H|(1-(\nabla I)H^{-1}(\nabla I)^T)}{|H|} = 1 - (\nabla I)H^{-1}(\nabla I)^T . \tag{13}$$

Similarly, $F_2$ has the following expression,

$$F_2 = \frac{|K_1|}{|H|} = \frac{|H|(1+(\nabla I)H^{-1}(\nabla I)^T)}{|H|} = 1 + (\nabla I)H^{-1}(\nabla I)^T \tag{14}$$

Since $F_1$ and $F_2$ are independent from the affine coefficient $P$, it is undoubted that $(\nabla I)H^{-1}(\nabla I)^T$ cannot be affected by the affine coefficients. This independence is critically important as considering the motion trend as the displacement goes to zero, $P: I_1 \to I_2$. Let $E$ represent the critically important term, involving the 1st- and 2nd-order derivatives:

$$E = (\nabla I)H^{-1}(\nabla I)^T \tag{15}$$

we have the following conclusion:

**Theorem 1**

$E = (\nabla I)H^{-1}(\nabla I)^T$ *is a differential affine invariant in the 3D Euclidean space.*

The proof of **Theorem 1** is given in the **Appendix**. From the proof of **Theorem 1,** we obtain the following relationships among the three invariants of $F_1$, $F_2$ and $E$:

$$\begin{cases} F_1 = 1 - E \\ F_2 = 1 + E \end{cases} \tag{16}$$

We notice that $F_1$ and $F_2$ are not independent and have the following special relationship,

$$F_1 + F_2 = 2 \tag{17}$$

Therefore, among the three invariants, $E$ should be the fundamental invariant because both $F_1$ and $F_2$ can be expressed by $E$. As a striking conclusion and an important property, $E$ will be used to construct the ESM feature maps and extract the elastic features as input to a classifier for CADx.

Mathematically, **Theorem 1** expands the scope of two widely used models of isometric deformation and conformal mapping, solving non-rigid deformation problems [16,22]. Besides their mathematical significance, $F_1$, $F_2$ and $E$ are all relevant to the elasticity of the soft tissues and have the following attractive properties and advantages over the traditional ESMs:

- *Informativeness*: Since the Harris operator is closely related to shape changes while the Hessian operator is associated with wave fluctuation, $F_1$, $F_2$ and $E$ contain rich elasticity information qualitatively and quantitatively related to the DEM.
- *Locality*: Since both Harris and Hessian operators are local tensors in the differential domain, $F_1$, $F_2$ and $E$ are all local scalars and describe the local properties of soft tissues' elasticity.
- *Generality*: The three invariants are all derived under the three very general assumptions made in section II.A and mimic most of the complex deformations related with the elasticity and wave fluctuation of the soft tissues. As such, they reflect the general measures of soft tissues' elasticity.
- *Invariance*: Invariants would be one of the most important properties of non-rigid shapes according to mathematical theory [16]. The three DEM descriptors of $F_1$, $F_2$ and $E$ are theoretically proved to be intrinsically invariant under very complex non-rigid shape deformations in the differential domain, indicating that they reflect the intrinsic property of the soft tissue's elasticity [23].
- *Simplification*: Since they are all differential invariants, the three DEM descriptors would be robust under non-rigid tissue deformations in theory. The elastic feature maps could be calculated from one image volume instead of several frames or multi-scanning. Therefore, the traditional mechanical vibration and tissue displacement may not be needed, which significantly simplify the process of elasticity computation.

*2.3. DEM-based Feature Maps and Feature Descriptors*

By consideration of their relationships among $F_1$, $F_2$ and $E$, we choose $E$ of **Eq. (15)** as our candidate to construct the elastic feature maps for representation of ESM (or DEM) of the soft tissues' elasticity. Theoretically, $E$ has several advantages as the representation of DEMs. To calculate the DEMs in a digital image, the first step is to calculate the 1st- and 2nd-order partial derivatives, see **Eq. (15)**. The difference of image voxel values is usually used for the derivative calculation. However, due to the noise in the image, a robust differential filter becomes a practical requirement. In this work, the Deriche filter is chosen to calculate the 2nd-order derivative [24], where the border size is 3, window size is 7×7×7 and alpha value is 1. For the same reason, the Sobel operator is chosen for the calculation of the 1st-order derivative [25].

Once $E$ is obtained, the next step is to generate its corresponding feature maps with digitalization of the elasticity measures. Practically, the value range of $E$ is very wide varying from $-2\times10^8$ to $2\times10^8$. To minimize information loss during the DEM digitalization, we devise a one-to-one mapping-based signal suppression method to produce the feature maps by:

$$Q = \begin{cases} \operatorname{atan}\left(\frac{1}{\sqrt[n]{E}}\right) & E \geq 0 \\ -\operatorname{atan}\left(\frac{1}{\sqrt[n]{|E|}}\right) & E < 0 \end{cases} \tag{18}$$

where $E$ of **Eq. (15)** is the local deformation invariant or DEM, $Q$ is the suppressed DEM, $n \in \mathbb{N}$ is called root power, and $\mathbb{N}$ is the associated integer set. The basic form of the suppression formula is plotted in **Fig. 2.**

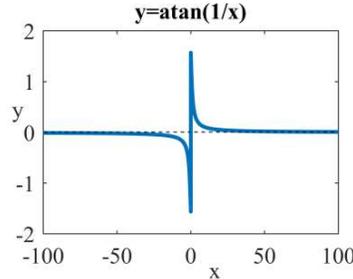

**Fig. 2**: The signal suppression function introduced by the 3D texture image generation, $y=\operatorname{atan}(1/x)$, where some very sharp signals are suppressed very close to 0 while some very weak but useful components are amplified to around $\pm \pi/2$.

Followed by the signal suppression via **Eq. (18)**, a linear gray scaling method is adopted to generate the suppressed DEM ($Q$) as shown in **Fig. 3**, where one representative slice is extracted from each 3D tumor to calculate and illustrate its suppressed DEM with different root powers where $n$ is set to be 1, 2, and 3. We only investigate the tumor in the region of interest (ROI). By this mapping method, the DEM of the two types of tumors' images, i.e. colon polyps and lung nodules, are extracted and demonstrated. Comparing with their original images, the suppression method can not only obtain striking pixel contrast but also produce more details with the growth order of $n$ as demonstrated by their histograms.

The one-to-one mapping-based signal suppression method demonstrates some very good characteristics in the generated 3D DEM feature maps as follows:

- Signal normalization: All elasticity measures represented by *E* with very wide scope can be mapped into a very narrow range using **Eq. (18)**. The function of suppression shows that it varies in $(-\infty, +\infty)$ while its output ranges from $-\frac{\pi}{2}$ to $\frac{\pi}{2}$. Therefore, **Eq. (18)** not only changes the distributions of the scalar to be more uniform and but also acts as a normalization mapping.
- Information completeness: Since this mapping of **Eq. (18)** is a one-to-one mapping, there is no information loss compared with the original signal after signal suppression.
- Noise suppression: The noise or very sharp signal can be successfully suppressed toward 0 after the mapping. Since the derivatives of noise images are always very sharp, which could bring severe challenges in signal processing [26], **Eq. (18)** suppresses sharp signals to a very low level and brings great benefit to the feature map generation.
- Weak signal amplification: From the suppression function of **Fig. 2**, it is seen that the mapping is also a very good amplifier, which can map some weak but useful signals (near the low-frequency area) to around $\pm \frac{\pi}{2}$ and produce more details. As a result, more micro-structures of tumors can be preserved, see **Fig. 3**.
- Contrast enhancement: The contrast of CT images can be enhanced to preserve more details hidden in the original images. This is an attractive point for low tissue contrast CT image analysis.

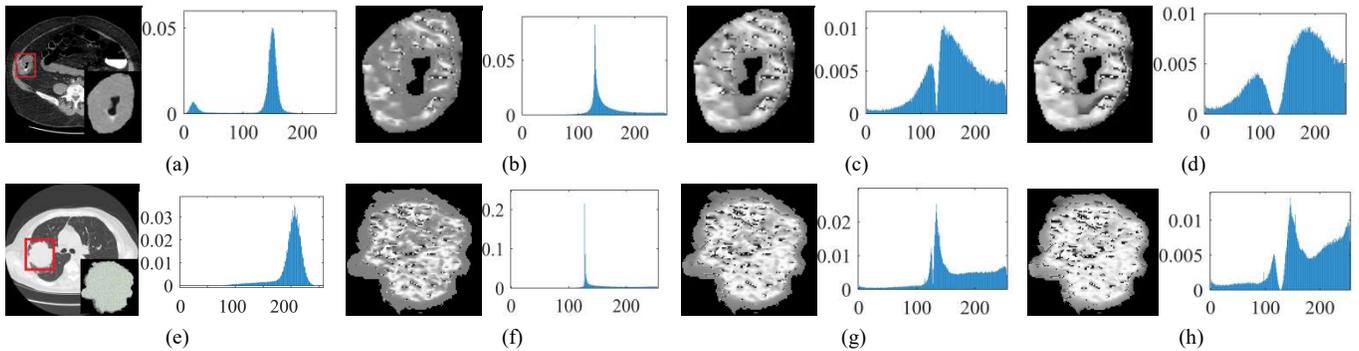

**Fig. 3**. Feature maps and their histograms of two tumors after DEM's suppression of **Fig. 2**, under three *n*-th roots where x-axis represents the bin of histogram, y-axis is the probability of every bin, and the number of gray levels is 256. (a) one original slice and its histogram in the 3D polyp volume; (b) *n*=1 (polyp); (c) *n*=2 (polyp); (d) *n*=3 (polyp); (e) one original slice and its histogram in the 3D lung nodule volume; (f) *n*=1 (nodule); (g) *n*=2 (nodule); (h) *n*=3 (nodule).

When elastic feature maps *Q* of **Eq. (18)** are generated as the DEM representation, the next step is to extract the elastic features from the maps to characterize a lesion. There are many methods available for this purpose, such as local binary pattern (LBP) [27], Weber local descriptor [28], Gabor [29], wavelet [30], gray level co-occurrence matrix (GLCM) [31], neighboring gray-level dependence matrix [32], gray level run-length matrix [33], and gray level size zone matrix [34]. In this study, we chose the GLCM as the feature matrix or surrogate from the elastic feature maps. This choice is a deliberate decision after full investigations on the GLCM's properties, such as 2D squared patterns, less influence from boundaries and postures, scaling resistance, and rotational robustness. In implementation, we chose the nearest 26 neighbors around the concerned voxel to determine the 13 independent directions to generate GLCMs. From each GLCM or along each of the 13 directions, 28 extended Haralick measures (eHMs) [35,36] are computed as the elastic feature descriptors, called DEM-based tumor descriptor (DEMTD), resulting in total 13×28=364 variables (i.e. 28 variables from each of the 13 directions).

## 3. Results

The above presented methods for DEMTD-based CADx were evaluated on two different lesion datasets: colon polyps and lung nodules. As mentioned earlier, the CT image of colon polyps can be taken in different body positions. The CT image of the lung nodule can be produced at any phase of one breath cycle. In both situations, the DEMTD is extracted from the images acquired at one position or one phase. For comparison purpose, the classification performance of DEMTD-based CADx was compared to several existing cML and DL methods using AUC (area under curve of the receiver operating characteristics – ROC) scores. In addition, accuracy (ACC), sensitivity (SN) and specificity (SP) are reported as references.

### 3.1 Datasets
#### 3.1.1 Colon polyp dataset

The patients, scheduled for clinical colonoscopy, were recruited to this study with CT colonography (CTC) scans under informed consent after approval by the Institutional Review Board (IRB). All polyps were subsequently removed. Total of 84 patient CTC scans with 87 polyps and their pathological reports were collected. Among the 84 patients, 51% are males and 49% are females with age ranges from 45 to 91 years old (about 66 years old on average). The pathological reports indicate two categories of malignant (adenocarcinoma) and benign (including serrated adenoma, tubular adenoma, tubulovillous adenoma and villous adenoma). The polyp dataset is balanced with 44 malignant and 43 benign. The polyp sizes range from 3 cm to 10 cm (mean of 4.5 cm).

*3.1.2    Lung nodule dataset*

The patients, scheduled for clinical CT-guided lung nodule needle biopsy, were recruited with IRB-approved informed consent. Total of 65 patient CT chest scans with 66 lung nodules and their pathological reports were collected. The average age of the patients is 69, ranging from 33 to 91 years old, and 52% of the patients are males and 48% are females. The lung dataset is unbalanced, including 46 malignant and 20 benign. The diameter of the nodules ranges from 9 mm to 130 mm (mean size of 32mm).

*3.1.3    Data preparation and evaluation setting*

For the above presented aML and reviewed cML, the contour of each polyp and lung nodule was drawn by an investigator in a slice-by-slice manner through the lesion volume inside the corresponding patient CT image using a semi-automated segmentation algorithm and confirmed by a radiologist to ensure accuracy. When computing the 1$^{st}$- and 2$^{nd}$-order derivatives for a voxel nearby the contour, the voxels outside the contour were included. The extracted features from the polyp and nodule volumes were the inputs to a CADx pipeline of aML or cML for feature selection and classification.

For the above-mentioned DL-based CADx, three CNN (convolutional neural network) architectures [37-39] were implemented. A rectangular volumetric mask was applied to all polyps/nodules to extract a volumetric ROI with the polyps/nodules located at the mask center. The ROIs of the lesions from the CT images were the inputs to the three CNN architectures.

Two-fold cross validation was applied for evaluation of the classification performances. For small datasets, two-fold cross-validation has equally distributed and maximized the use of the datasets. From the polyp database, we randomly selected 21 samples from the benign group and 22 from the malignant group for training. The remaining polyps were used for testing. Similarly, 10 benign and 23 malignant lung nodules were randomly chosen for training and the remaining samples were used for testing. For each database, the experiment of randomly selecting the training and testing were repeated 50 times. The results are the average of 50 two-fold cross validation runs and the standard deviation of the 50 runs.

*3.2    Experimental Outcomes*

*3.2.1    DEM based GLCM calculation and illustration*

Before GLCM calculation, the elastic feature maps are computed by **Q** of **Eq. (18)**, where the root power was set by varying from 1 to 9 and **E** is our DEM representation of lesion features calculated by **Eq. (15)**. When calculating GLCM from the computed 3D feature maps, there are several important parameters such as directions and gray levels [31]. As mentioned before, considering the 26 nearest neighbors, a total of 13 independent directions were selected [35]. The number of gray levels was determined by a trial-and-error manner among 16, 24, 32, 40, 48, 56, 64, 72, 80, 88, 96, 104, 112, 120 and 128. More details on the determination of the root power and the gray levels will be given in the next section below.

To illustrate the DEM-based GLCMs at different gray levels, we plotted the GLCM patterns at two gray levels of 48 and 80, respectively, as shown in **Fig. 4**. This figure directly shows that different lesions have distinct patterns expressed by the DEM-based GLCMs, which visually demonstrate the potential of DEM representation as elastic features for characterization of lesions. In addition, the DEM-based GLCMs at different directions show different patterns and, thus, bring more discriminative features for lesion classification.

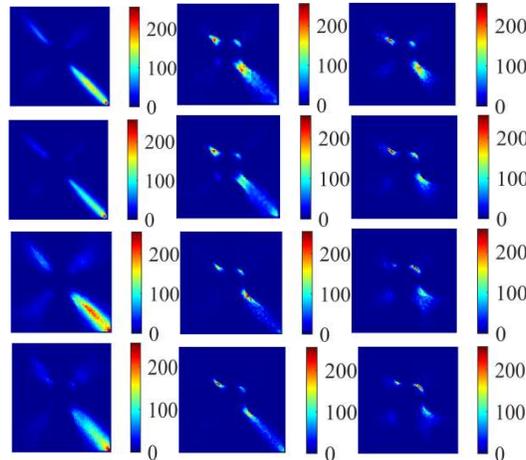

**Fig. 4**. The GLCM patterns in DEM-based lesion representations extracted from three different polyps (column) with two directions and two gray levels. The GLCMs in the 1$^{st}$ and 2$^{nd}$ rows are calculated along the direction (1,0,0) with 48 and 80 gray levels respectively. The 3$^{rd}$ and 4$^{th}$ rows illustrate the GLCMs of the direction (0,1,1). Their gray levels are 48 and 80 as well.

*3.2.2    Classification performances*

As mentioned above, total of 364 elastic feature measures were computed from the 13 DEM-based GLCMs of each polyp or nodule as input to a classifier. Because these elastic feature measures are independent scalers without any logical or topological relevance, it is hard to find a reasonable permutation for them. Moreover, different permutations could generate different convolutional results after convolution performances [40]. Based on our experiences [36], the random forest (RF) classifier has some advantages in the permutation. It uses random sampling in the procedure of tree generation via randomly splitting and combining the measures. It chooses thresholds in a non-uniform or adaptive manner. Therefore, the RF package [41] was selected as the classifier for our DEM-based CADx pipeline. In this study, we used a function of the R-package "randomForest" to construct the "forest" and selected the importance order on the tree notes. An average of the total decrease in node impurities over all trees was computed, which is measured by the GINI impurity. We shared the same parameters for other arguments such as $\sqrt{13 \times 28} \approx 19$ variables for m-try at each node and 5000 trees in the forest. To address the redundancy among the features, we also used the forward-step feature selection (FSFS) method [42] to generate the best performance as the final tumor classification result.

As mentioned above, among all parameters in the calculation of DEM-based GLCMs, there are two most critical ones: the root power and the gray levels. Their different combinations would produce many groups of DEM-based GLCMs, resulting in many different elastic feature descriptors. As we mentioned in the previous section, the root power of $E$ is varying from 1 to 9 and the number of gray levels would be varying from 16 to 128 in an interval of 8. Thus, where would be 9×16=144 groups of elastic feature descriptor set for the polyps and nodules respectively. For each elastic feature descriptor set of 364 variables, we fed them to the FSFS method in the pipeline of RF to perform classification. By consideration of the complex relationship between the AUC score and the two parameters, it is a challenge to search their combination for the best AUC score. Thereafter, we use a brute force method to search that combination from the 144 groups of the descriptor set. After classification on the polyp dataset, we found that the best classification performance is obtained by 6 variables out of the total of 364 ones and reaches the AUC score of 0.942±0.031 when the $n$-th root power is 4 and number of gray levels of the DEM-based GLCM is 104.

Because the nodule dataset is not only severely unbalanced but also has more complex and large variation, including size and shape, the established Karhunen-Loève (KL) transform was applied along the 13 directions to relieve the correlation among the GLCM measures. The KL transformed GLCM measures were chosen as the elastic feature descriptors and fed to the FSFS method in the pipeline of the RF to perform the classification. The RF classifier predicted the best AUC score of 0.874±0.046 when the nth-root power is set to 8 and the number of gray levels of the DEM-based GLCM is 128. It shall be noted that the nodule dataset includes only the undetermined cases after experts' assessment on the nodules and these nodules were recommended for CT-guided needle biopsy. In other words, the experts' score on the nodules would be around 0.5. The predicted AUC score over 0.8 is an outstanding performance.

*3.2.3 Performance comparisons with several existing CADx methods*

Total ten existing ML (including cML and DL) methods were implemented to perform comparison studies, including:

- VGG16 – It is a CNN benchmark model which has deep design for automatic feature extraction [37]. We selected the VGG16-layer model (the configuration D) with weights pre-trained on the ImageNet.
- ResNet18 – It is another CNN benchmark with very deep design and residual design for feature extraction [38]. The model is pre-trained on the ImageNet and fine-tuned by our two lesion image datasets.
- InceptionV3 – This ImageNet-pertained model is also a CNN benchmark with deep design and Inception block to capture more details at variable scales. The model is further fine-tuned by our two lesion image datasets [39].
- Intensity image based extended Haralick features (eHF) – It is a typical feature extraction method including total of 56 variables (or image features) extracted from the intensity image GLCMs over the 13 independent directions [31,35].
- Intensity image based extended Haralick measures (called eHM) and KL-transformed measures over eHM (called eHM+KL) – Both methods include all the 364 measures computed from the intensity image GLCMs over the 13 directions [35].
- CoLIAGe – It considers only the gradient angles in the 1$^{st}$-order derivative domain and extracts the entropy of every local patch to form a global texture descriptor by two joint histograms [43].
- Affine gradient LBP (AGLBP) – It investigates affine invariant by combination with traditional LBP to construct an affine robust descriptor to solve the vision problem [44].
- Affine differential local mean ZigZag pattern (ADLMZP) – It proposes a zigzag pattern and combines affine invariant to form a new descriptor to mitigate the affine challenges in computer vision application [45].
- DNN – It is a deep neuron network (4 layers, each with 500, 800, 1000, 2 neurons with ReLU and softmax as activations) [46]. It was trained on the extracted measures from the DEM-based GLCMs over the 13 directions.

For the three CNN-based DL architectures, the center or largest area slice was selected for classification. The single-channel CT was converted to 3 sub-channels by threshold [-600, 200], [-160, 600], and [-100, 200], as low, high and medium attenuation. Then all the inputs were resized to the same size (224×224 for VGG16 and ResNet; 299×299 for InceptionV3). The pre-trained models were fine-tuned for 10 iterations with stochastic gradient descent and learning rate 0.001. Cross-Entropy loss was adapted.

In the following six ML methods, i.e. eHF, eHM, eHM+KL, CoLIAGe, AGLBP and ADLMZP, we used the RF as the classifier with the same parameters as mentioned before. To see the different performance of RF classifier via DNN classifier, we compared both RF and DNN classifications on the same DEMTD features set. **Table 1** and **Table 2** list their top classification performances on the two pathologically proven datasets respectively.

For the polyp dataset, see **Table 1**, the three CNN-based DL methods generated comparable results as our previous CNN-DL investigations [47], i.e. CNN-DL with inputs of 2D image slices generated AUC scores in the range from 0.60 to 0.83, and CNN-DL with inputs of 3D image volumes generated AUC scores in the range from 0.80 to 0.84. The three intensity image-based GLCM Haralick texture measures, eHF, eHM and eHM_KL, consider the lesion texture properties and, therefore, improve the lesion classification. The angular-based GLCM model of CoLIAGe did not show advantages compared to the intensity image-based GLCM methods. Both the Affine gradient local pattern AGLBP and the Affine differential local mean pattern ADLMZP measures performed similarly to the intensity image-based GLCM measures. The presented tissue elasticity texture descriptor or DEMTD with RF classification improved the AUC score compared to other image feature extraction methods. It is noted that the DNN as a classifier on the same feature set (DEMTD+DNN) did not show advantage over the RF classifier on the same feature set.

TABLE 1: COMPARISON OF CLASSIFICATION PERFORMANCES OF OUR METHOD (DEMTD+RF) AND THE 10 EXISTING METHODS ON THE POLYP DATASET

| Method | AUC | ACC | SN | SP |
|---|---|---|---|---|
| VGG16 | 0.788 | 0.788 | 0.790 | 0.786 |
| ResNet18 | 0.665 | 0.674 | 0.699 | 0.652 |
| InceptionV3 | 0.547 | 0.609 | 0.583 | 0.631 |
| eHF+RF | 0.884 | 0.840 | 0.829 | 0.850 |
| eHM+RF | 0.895 | 0.843 | 0.812 | 0.874 |
| eHM+KL+RF | 0.909 | 0.854 | 0.817 | 0.891 |
| CoLIAGe+RF | 0.699 | 0.707 | 0.711 | 0.702 |
| AGLBP+RF | 0.858 | 0.818 | 0.800 | 0.835 |
| ADLMZP+RF | 0.880 | 0.840 | 0.825 | 0.855 |
| DEMTD+DNN | 0.839 | 0.830 | 0.793 | 0.867 |
| DEMTD+RF | 0.942 | 0.907 | 0.912 | 0.902 |

For the nodule dataset, see **Table 2**, most of the implemented CADx methods performed relatively similar as for the polyp dataset, except for the overfitting problem for the ResNet-18, Inception-V3, and DNN. Some possible reasons may be due to the huge variations of the nodules and the data imbalance. Therefore, their results are not reported in **Table 2**.

TABLE 2: COMPARISON OF CLASSIFICATION PERFORMANCES OF OUR METHOD (DEMTD+KL+RF) AND THE 7 EXISTING METHODS ON THE LUNG NODULE DATASET

| Method | AUC | ACC | SN | SP |
|---|---|---|---|---|
| VGG16 | 0.669 | 0.687 | 0.673 | 0.700 |
| eHF+RF | 0.647 | 0.734 | 0.897 | 0.360 |
| eHM+RF | 0.719 | 0.747 | 0.899 | 0.397 |
| eHM+KL+RF | 0.792 | 0.805 | 0.881 | 0.631 |
| CoLIAGe+RF | 0.654 | 0.756 | 0.930 | 0.358 |
| AGLBP+RF | 0.736 | 0.774 | 0.875 | 0.540 |
| ADLMZP+RF | 0.782 | 0.809 | 0.879 | 0.650 |
| DEMTD+KL+RF | 0.874 | 0.860 | 0.887 | 0.804 |

It is very interesting to see the lesion classification performances of all the implemented CADx methods in terms of quantitative measures of AUC, ACC, SN, and SP. The proposed DEMTD with RF classification has not only the highest AUC scores but also the most consistent AUC, ACC, SN and SP for both the pathologically proven image datasets of limited sizes. The AUC stands out as the most important figure-of-merits (FOM) among the four measures. The AUC score on the polyp dataset increased by 5%-10% and on the nodule dataset by 9%-20% by the proposed model-based feature extraction. To verify the significant difference of the AUC gain, the T-Tests were performed as the significant test based on the predict scores to conduct comparisons as shown in **Table 3** where all p-values indicate that the proposed tissue elastic feature extraction model is significantly different from the other state-of-the-art CADx methods.

TABLE 3: T-TEST RESULTS OVER PREDICTION SCORES BETWEEN THE PRESENTED METHOD AND THE EXISTING METHODS

| Dataset \ Methods | VGG16 | ResNet18 | InceptionV3 | eHF+RF | eHM+RF | eHM+KL+RF | CoLIAGe+RF | AGLBP+RF | ADLMZP+RF | DEMTD+DNN |
|---|---|---|---|---|---|---|---|---|---|---|
| Polyp | <<0.05 | <<0.05 | <<0.05 | <<0.05 | 0.019 | 0.024 | <<0.05 | <<0.05 | 0.009 | <<0.05 |
| Nodule | <<0.05 | -- | -- | <<0.05 | <<0.05 | 0.016 | <<0.05 | <<0.05 | 0.013 | -- |

4. Conclusion and Discussions

In medical diagnosis of many diseases, we frequently encounter the challenges of having limited data to achieve the diagnoses we wanted. Differentiation of the malignant from benign colon polyps and lung nodules is a typical example. Experts have shown great intelligence in incorporation of previous knowledge about the lesion properties and integration of the prior knowledge with the limited data to achieve the tasks they desired. The possibility of training machines to function in a manner similar to the human intelligence is known as artificial intelligence (AI) and, thus, the above presented model-based feature extraction is also called AI-driven feature extraction hereafter. This study explored the potential of the AI-driven feature extraction for prediction of lesion malignancy and demonstrated its outstanding performance in differentiating malignance from benign lesions using two pathologically proven CT image datasets of colon polyps and lung nodules of limited sample size. Because of its modeling of the tissue elastic properties in extraction of the corresponding tissue discriminative features from the acquired images for machine leaning (or ML), the gain by the presented adaptive ML (or aML) over the conventional ML (or cML) is expected.

The main reason why the deep learning (or DL) did not perform very well compared to both the aML and cML may be due to the small sample size (<85). Another reason may be due to the format of the input data. Both the aML and cML input the features in a unform format which are derived from the CT images including only the lesion volumes, but the DL input the CT images which have variable sizes depending on the lesion sizes. Other reasons, such as large variations in CNN model designs and model parameter settings, can be seen in the studies of DL in medical imaging-based diagnosis [48,49].

One way to mitigate the effect of the variation in input data format to DL may be taking the GLCMs as the input data to the DL [47], because the GLCMs derived from both the original CT images of the lesion volumes and the DEM-based elastic image maps $E$ or **Eq. (15)** of the lesion volumes, are all normalized data and share the same size. To expand this GLCM-DL approach to include the modeling of the tissue properties is our next research topic beyond the scope of this study. A more attractive research topic to us would be to include the AI modeling strategy into the DL architecture design for simultaneous feature extraction and classification.

One limitation of this study is related to the presence of singular points of the Hessian matrices. Currently, we just set their DEMs values to be zero, which is just a stopgap, not an ideal solution. Another limitation of this study is the very small sample size of the two medical diagnosis datasets. Validating this proposed method to large datasets by a multi-institutional effort will be one of our future research topics. Lastly, we note the absence of a comparison with a relevant method, which directly measures the tissue elasticity by acquiring a series of images with certain mechanical vibration equipment as the Elastographic Imagers do [5-7].

**Appendix**

Here we supply the proofs for the property, lemma, and theorem discussed in the main text.

**Property 1**
*Harris operator, $G$, after a local affine transformation keeps*
$$\mathcal{G} = P^T G P$$
*where $\mathcal{G}$ is the subsequent Harris operator, $P$ and $P^T$ are the local affine projection and its transpose.*
**Proof:**
After affine transformation, the 1$^{st}$ order partial derivatives will be expressed as:
$$I'_X = \frac{\partial I}{\partial X} = \frac{\partial I}{\partial x}\frac{\partial x}{\partial X} + \frac{\partial I}{\partial y}\frac{\partial y}{\partial X} + \frac{\partial I}{\partial z}\frac{\partial z}{\partial X} = p_{11}I'_x + p_{21}I'_y + p_{31}I'_z$$
$$I'_Y = \frac{\partial I}{\partial Y} = \frac{\partial I}{\partial x}\frac{\partial x}{\partial Y} + \frac{\partial I}{\partial y}\frac{\partial y}{\partial Y} + \frac{\partial I}{\partial z}\frac{\partial z}{\partial Y} = p_{12}I'_x + p_{22}I'_y + p_{32}I'_z$$
$$I'_Z = \frac{\partial I}{\partial Z} = \frac{\partial I}{\partial x}\frac{\partial x}{\partial Z} + \frac{\partial I}{\partial y}\frac{\partial y}{\partial Z} + \frac{\partial I}{\partial z}\frac{\partial z}{\partial Z} = p_{13}I'_x + p_{23}I'_y + p_{33}I'_z$$
and we get the new Harris operator:
$$\mathcal{G} = \begin{bmatrix} I'_X I'_X & I'_X I'_Y & I'_X I'_Z \\ I'_X I'_Y & I'_Y I'_Y & I'_Y I'_Z \\ I'_X I'_Z & I'_Y I'_Z & I'_Z I'_Z \end{bmatrix} = \begin{bmatrix} p_{11} & p_{21} & p_{31} \\ p_{12} & p_{22} & p_{32} \\ p_{13} & p_{23} & p_{33} \end{bmatrix} * \begin{bmatrix} I'_x I'_x & I'_x I'_y & I'_x I'_z \\ I'_x I'_y & I'_y I'_y & I'_y I'_z \\ I'_x I'_z & I'_y I'_z & I'_z I'_z \end{bmatrix}$$
$$* \begin{bmatrix} p_{11} & p_{12} & p_{13} \\ p_{21} & p_{22} & p_{23} \\ p_{31} & p_{32} & p_{33} \end{bmatrix} = P^T G P$$

**Lemma 1**
*$F_1$ and $F_2$ are independent of the local affine mapping $P$, and they are two differential affine invariants in 3D Euclidean space.*
**Proof:**
sBased on **Eq. (9)** and **Eq. (10)**, we have:
$$\mathcal{F}_1 = \frac{|\mathcal{K}_1|}{|\mathcal{H}|} = \frac{|\mathcal{H} - \mathcal{G}|}{|\mathcal{H}|} = \frac{|P^T H P - P^T G P|}{|P^T H P|} = \frac{|P^T (H - G) P|}{|P^T H P|}$$
$$= \frac{|P^T||H - G||P|}{|P^T||H||P|} = \frac{|H - G|}{|H|} = \frac{|K_1|}{|H|} = F_1$$
where $\mathcal{F}_1$, $\mathcal{K}_1$ are the subsequent $F_1$ and $K_1$. The proof of $F_2$ is similar to $F_1$. The mathematical derivation of **Eq. (11)** is shown as the following:
$$|K_1| - |H| = \begin{vmatrix} I''_{xx} - I'_x I'_x & I''_{xy} - I'_x I'_y & I''_{xz} - I'_x I'_z \\ I''_{xy} - I'_x I'_y & I''_{yy} - I'_y I'_y & I''_{yz} - I'_y I'_z \\ I''_{xz} - I'_x I'_z & I''_{yz} - I'_y I'_z & I''_{zz} - I'_z I'_z \end{vmatrix} - \begin{bmatrix} I''_{xx} & I''_{xy} & I''_{xz} \\ I''_{xy} & I''_{yy} & I''_{yz} \\ I''_{xz} & I''_{yz} & I''_{zz} \end{bmatrix} =$$
$$-(I'_y I'_y I''_{xx} I''_{zz} + I'_x I'_x I''_{yy} I''_{zz} + I'_z I'_z I''_{xx} I''_{yy} + 2 I'_y I'_z I''_{xy} I''_{xz}$$
$$+ 2 I'_x I'_y I''_{xz} I''_{yz} + 2 I'_x I'_z I''_{xy} I''_{yz} - 2 I'_x I'_z I''_{xz} I''_{yy} - I'_y I'_y I''_{xz} I''_{xz}$$
$$- 2 I'_y I'_z I''_{xx} I''_{yz} - I'_x I'_x I''_{yz} I''_{yz} - 2 I'_x I'_y I''_{xy} I''_{zz} - I'_z I'_z I''_{xy} I''_{xy})$$
$$= -(I'_x \quad I'_y \quad I'_z) H^*(I'_x \quad I'_y \quad I'_z)^T = -(\nabla I) H^* (\nabla I)^T$$
where $H^*$ is the adjoint matrix of $H$, and $H^* = |H| H^{-1}$. Then we have:
$$|K_1| - |H| == -(\nabla I) H^* (\nabla I)^T = -(\nabla I)|H|H^{-1}(\nabla I)^T = -|H|(\nabla I) H^{-1} (\nabla I)^T$$

**Theorem 1**
$E = (\nabla I) H^{-1} (\nabla I)^T$ *is a differential affine invariant in 3D Euclidean space.*
**Proof:**
According to **Eq. (11)**, we have
$$|K_1| - |H| == -|H|(\nabla I) H^{-1} (\nabla I)^T$$
where $H^{-1}$ is the inverse matrix of $H$. Then we have:
$$|K_1| = |H| - |H|(\nabla I) H^{-1} (\nabla I)^T = |H|(1 - (\nabla I) H^{-1} (\nabla I)^T)$$
So, we have:
$$F_1 = \frac{|K_1|}{|H|} = \frac{|H|(1 - (\nabla I) H^{-1} (\nabla I)^T)}{|H|} = 1 - (1 + (\nabla I) H^{-1} (\nabla I)^T)$$
$$= 1 - E$$
Similarly, we get:
$$F_2 = \frac{|K_2|}{|H|} = \frac{|H|(1 + (\nabla I) H^{-1} (\nabla I)^T)}{|H|} = 1 + (\nabla I) H^{-1} (\nabla I)^T$$
$$= 1 + E$$
Since $F_1$ and $F_2$ are local invariants, $(\nabla I) H^{-1} (\nabla I)^T$ must be invariant.

## Acknowledgments

This work was partially supported by the NIH/NCI grant #CA206171 and #CA220004.

## Conflict of Interest

The authors declare that they have no known conflicting financial interests or personal relationships that could have appeared to influence the work reported in this paper.

## Data Availability

The data are not publicly available due to privacy or ethical restrictions. Data can be made available upon request to the corresponding author.